**Editorial**

# Special Issue on the Atomic and Molecular Processes in the Ultracold Regime, the Chemical Regime, and Astrophysics

Guest Editors
**James F Babb** Harvard-Smithsonian Center for Astrophysics, USA
**Robin Côté** University of Connecticut, USA
**Hossein R Sadeghpour** Harvard-Smithsonian Center for Astrophysics, USA
**Phillip C Stancil** University of Georgia, USA

**Scope**

Professor Alexander Dalgarno (1928-2015) was a true giant in the field of atomic and molecular physics whose pioneering research spanned more than six decades. He not only had a lasting impact, but often led the field and set directions for research in the physics of the upper atmosphere (aeronomy), atomic and molecular astrophysics, astrochemistry, atomic and molecular collisions and interactions, ultracold collisions and ultracold chemistry, as well as areas which critically depend on atomic and molecular physics. He is regarded as the father of molecular astrophysics and of ultracold chemistry. His genius was in recognizing that physics was about solutions to its fundamental equations and he therefore developed methods which readily, even after several decades, found continued applications in a wide range of topics.

Much of what he accomplished fell squarely in the purview of this journal of record in atomic and molecular physics. Indeed, Professor Dalgarno published 22 papers in J. Phys. B from its start in 1968. We believe that the present Special Issue—devoted to original research topics close to Alex's heart and contributed by his many collaborators and students—is a fitting tribute. Below, we discuss in brief each of the papers in this Special Issue and provide context to, and contact with, the science of Alex Dalgarno[i].

**Short-range photoassociation from the inner wall of the lowest triplet potential of $^{85}Rb_2$**
R A Carollo, J L Carini, E E Eyler, P L Gould, and W C Stwalley

Studies of bound-free (and free-bound) transitions in molecular systems have been a staple of computational molecular physics for many decades including applications to radiative association and photodissociation processes. Dalgarno and co-workers have led the community on such calculations and were also at the forefront when laser-induced photoassociation was devised as a technique to create ultracold molecules in the laboratory. In their manuscript entitled ``Short-range photoassociation from the inner wall of the lowest triplet potential of $^{85}Rb_2$'' Carollo et al. present a combined experimental-theoretical investigation of photoassociation. However, instead of focusing on the more typical long-range region, they explore photoassociation at short internuclear distances, near the inner turning-point of the incoming channel. Further, the model for this mechanism considered in their paper is based on work 46 years ago by Allison and Dalgarno (1971), who demonstrated the continuity of the absorption (or emission) cross section as the initial wave function passed through the dissociation limit, i.e., evolving from a bound-bound to a bound-free process. In the current manuscript, the cross sections become appreciable when the square of the initial state wave function has a local maximum near the inner turning point. Comparisons to experimental data for $^{85}$Rb$_2$ show reasonable agreement with the model.

## Cold NH-NH collisions in a magnetic field: Basis set convergence vs sensitivity to the interaction potential

Yu V Suleimanov and Timur V Tscherbul

The theory for inelastic transitions of a rigid-rotor due to heavy-particle collisions was formulated in the seminal work of Arthurs and Dalgarno (1960) which laid the foundation for related calculations up until the present day. The approach has been utilized in countless studies and motivated the development of numerous inelastic scattering computational packages. Dalgarno and co-workers extended the method to include transitions amongst molecular fine-structure levels and the effects of an external magnetic field (Krems and Dalgarno 2004). Here, Suleimanov and Tscherbul apply the method to spin-polarized NH molecules as detailed in their manuscript "Cold NH-NH collisions in a magnetic field: Basis set convergence versus sensitivity to the interaction potential." They explore the effects of basis set size and the accuracy of the potential energy surface on the resulting elastic and inelastic cross sections. Interestingly, variation of the potential has less influence on the results compared to basis set truncation, which can affect the width of resonances in the cold regime.

## Radiative charge transfer in collisions of C with He$^+$

James F Babb and B M McLaughlin

Charge transfer has been known to play an important role in the elemental ionization balance of a variety of astrophysical environments since the 1970s (Field and Steigman 1971). Such processes are usually driven by nonadiabatic interactions. However, for singly-charged systems, nonadiabatic couplings are often weak so that radiative charge transfer dominates. In this process, a photon is emitted during the interaction transitioning the system from an excited electronic molecular state to a lower state. Dalgarno and co-workers pioneered radiative charge transfer calculations more than 50 years ago (Allison and Dalgarno 1965) and routinely performed such computations over the years. The suggestion (Côté and Dalgarno 2000) more than 15 years ago that radiative charge transfer is important in the cold to ultracold regimes inspired many relevant calculations and measurements, e.g., Rellergert *et al.* (2011), making this an active field today. In their manuscript, ``Radiative charge transfer in collisions of C with He$^+$,'' Babb and McLaughlin revisit a system studied by Dalgarno nearly 25 years ago. Here they update the molecular structure using modern quantum chemistry methods, while an optical potential approach and the distorted-wave approximation are used for the dynamics.

## Resonances at very low temperature for the reaction D$_2$ + H

I. Simbotin and R. Côté

Trapped cold molecules offer a range of possible applications in chemical reactivity and high precision metrology, which go beyond those already implemented for atoms because molecules have additional degrees of freedom for excitation and population (vibration and rotation). These same virtues also make (most) molecules not readily amenable as workhorses for cooling and trapping techniques used for atoms, such as laser cooling and Zeeman slowing. Nevertheless, investigations of diatom-atom collisions have given considerable insight into properties of cold molecular systems. In the 1990s, Sun and Dalgarno (1994) and Balakrishnan et al. (1997a,b) performed accurate reactive and inelastic scattering calculations, respectively, for $H_2 + H$, extending down to the ultracold regime. In their manuscript, Simbotin and Côté study an isotopologue of this benchmark collision system augmenting their earlier work on $H_2 + D$ to new computations for $D_2 + H$. Using a hyperspherical approach on a modern $H_3$ potential energy surface, they obtain rovibrationally-resolved rate coefficients for reactive and inelastic scattering from the ultracold to the thermal regime, taking into account the nuclear spin symmetry of $D_2$. As might be expected, they find significant isotopic variation in the resonance structures.

**[Laser slowing of CaF molecules to near the capture velocity of a molecular MOT](#)**
Boerge Hemmerling, Eunmi Chae, Aakash Ravi, Loic Anderegg, Garrett K Drayna, Nicholas R Hutzler, Alejandra L Collopy, Jun Ye, Wolfgang Ketterle, and John M Doyle

To obtain high numbers of trapped molecules, upon loading into a MOT, a range of alternative and complementary techniques have been tried and proposed. In this work, CaF, one of the first molecules to be trapped, is chosen due to its light mass, large electric dipole moment, and favorable optically closed vibrational A - X transition rates. The molecules are cooled in collision with He buffer gas, are allowed to diffuse in free space for 30 cm, and in a short (20 cm) distance to the detection plates, are cooled by a counter-propagating laser beam. This technique allows for a large number of molecules (60,000) to reach the MOT capture volume by increasing the solid angle available for capture.

**[Radiation pressure force from optical cycling on a polyatomic molecule](#)**
Ivan Kozyryev, Louis Baum, Kyle Matsuda, Boerge Hemmerling and John M Doyle

For molecules larger and more complex than diatomics, more innovation is necessary to achieve laser cooling. Stark deceleration of dipolar molecules, in successive and alternating electric field regions, has been used to slow molecules down to about 1 K. Laser cooling of closed vibrational transitions is made more complicated because even in vibrational motion polyatomic molecules can bend, in addition to stretching, their bonds. Here, radiation pressure due to two-color deflecting lasers, in perpendicular direction to the He cooled SrOH molecular beam, is used to slow molecules. An average of 100 photons scatter on a closed rotational transition. The spatial profile of the slowed molecules is imaged via laser-induced fluorescence.

**[(Letter) Mapping trilobite state signatures in atomic hydrogen](#)**
Jesús Pérez-Ríos, Matthew T Eiles and Chris H Greene

Line shift and broadening are quintessential hallmarks of interaction in a medium. It was the Rydberg line shift measurements of Amaldi and Segre (1934) which provided the clue for Fermi (1934) to develop his mean field s-wave scattering pseudopotential theory, now better known as the Fermi contact interaction. In ultracold systems, this same interaction leads to formation of an exotic class of molecules, known as ultralong range Rydberg molecules. In this work, the authors combine this theory with the static binary line broadening theory to obtain line shape profiles in high density collision of H-H(*nl*). The electron scattering length from H(1*s*) is positive for both total electron spin states, but because the Born-Oppenheimer potential energy curves still exhibit the underlying Rydberg electron oscillations, the authors nevertheless unearth sharp line shapes which depend on the degree of Rydberg excitation, but not on the perturber gas density.

**Efficient non-parametric fitting of potential energy surfaces for polyatomic molecules with Gaussian processes**
Jie Cui and Roman V Krems

In studies of the potential energy surfaces of polyatomic molecules, even after large-scale on-grid computations of multi-dimensional Born-Oppenheimer energy values, the daunting non-linear task of multi-surface fitting to analytic multi-mode functions remains. A whole class of literature in molecular physics and chemistry is devoted to the topic. In this work, machine-learning techniques are used to implement a Gaussian regression method efficiently to fit multi-dimensional molecular surfaces. In one example, it is shown that a potential surface for the $N_2$ dimer is reproduced, at the same level of accuracy, with one order of magnitude less energy point calculations.

**(Letter) First prediction of inter-Coulombic decay of $C_{60}$ inner vacancies through the continuum of confined atoms**
Ruma De, Maia Magrakvelidze, Mohamed E Madjet, Steven T Manson and Himadri S Chakraborty

Photoionization is a basic process that can be used as a powerful probe of the electronic structure of an atom or molecule. Dalgarno devoted many works to calculations of photoionization processes for atoms and molecules, mainly diatoms---though we note that one of his first papers was on the photoionization of methane (1952), and to their applications in atmospheric physics and astrophysics. Chu and Dalgarno (2004) explored time-dependent density functional theory (TDDFT) to calculate polarizabilities and van der Waals coefficients for a large fraction of the atoms in the periodic table.
In their paper, "First prediction …", De *et al.*, study the photoionization of Ar and Kr atoms confined in $C_{60}$ molecules (endofullerenes). Using TDDFT calculations to model the dipole response of the $C_{60}$ molecules to externally applied photons, they demonstrate the existence of distinctive resonances that are signatures of resonant inter-Coulombic decay in these systems.

**Modelling the role of electron attachment rates on column density ratios for $C_nH^-/C_nH$ (n = 4; 6; 8) in dense molecular clouds**
F A Gianturco, T Grassi and R Wester

In their paper "Modelling the role …", Gianturco *et al.* present detailed models for the abundances of $C_nH$ (n=4, 6, and 8) and their anions in molecular clouds. As they note, the identification of molecular anions in astrophysical spectra is relatively recent, while the formation of molecular anions by radiative attachment was investigated theoretically some time ago by Dalgarno and McCray (1973) and by Herbst (1981). In later studies, Dalgarno and collaborators extended the investigations of attachment to the formation of negative ions of polycyclic aromatic hydrocarbons in diffuse (Lepp and Dalgarno 1988a) and dense clouds (Lepp *et al.* 1988b) and Dalgarno highlighted the likelihood of the formation of carbon chain anions (2000). It was many years later that spectra observed in the molecular cloud TMC-1 were identified as $C_6H^-$ (McCarthy et al 2006). Now there are several more observational confirmations in other environments of the ISM, but the need persists for better descriptions of the chemistry. In their paper, Gianturco *et al.* explore quantitative effects of their previously calculated rate coefficients governing anion astrochemistry.

**Geometric phase effects in ultracold hydrogen exchange reaction**
Jisha Hazra, Brian K Kendrick and N Balakrishnan

Problems in the scattering of atoms and diatoms are at the heart of reactive chemical dynamics, while the cross sections and rate coefficients are of great importance to astrophysical and astrochemical modeling, particularly when H is involved. Dalgarno and collaborators used semi-classical and quantum methods to study reactions at thermal and ultracold temperatures. In earlier work, Dalgarno and McCarroll (1956-7) investigated the treatment of the coupling between electronic states, due to the nuclear kinetic energy, beyond the Born-Oppenheimer approximation. Later, Zygelman (1990) related such couplings to the presence of a non-Abelian gauge potential in certain long-range interactions. Conical intersections in potential energy surfaces governing certain atom-diatom collisions are also related to non-Abelian geometric phases. In their paper "Geometric phase effects…", Hazra *et al.* apply quantum reactive scattering techniques to hydrogen exchange scattering systems with conical intersections, where geometric phase effects are expected to occur in the collisional dynamics. By isolating specific rotational channels, they show that such effects might be observable at low temperatures.

**Oscillator strengths for $1s^2\ ^1S_0$ - $1s2p\ ^3P_{1;2}$ transitions in helium-like carbon, nitrogen and oxygen including the effects of a finite nuclear mass**
Donald C Morton and G W F Drake

Pioneering work by Drake and Dalgarno (1969) showed how variationally optimized basis sets could be combined with fine structure effects to predict reliable radiative properties of intercombination lines in helium-like ions, which appear in astrophysical observations of ionized environments, such as coronae and X-ray emitting plasmas. Drake and collaborators have extensively developed basis set methods for precision calculations of few-electron systems, which continue to find applications. For example, theoretically determined isotope shifts of the $^7Be^+$ ion (Yan, Drake, and Nörtershäuser 2008, 2009) were used to place constraints on the production of $^6Li$ in novae using astronomical observations of resonance lines (Tajitsu *et al.* 2015). In their contribution, Morton and Drake "Oscillator strengths …" present improved calculations of radiative properties for highly-charged helium-like ions of C, N, and O, including the effects of finite nuclear masses.

**Exploration of the origin of anomalous dependence for near-threshold harmonics in $H_2^+$ on the ellipticity of driving laser fields**
K Nasiri Avanaki, Dmitry A Telnov and Shih-I Chu

A benchmark molecular ion for theoretical physics and chemistry, $H_2^+$ is also a fundamental species in astrophysical environments. Dalgarno and co-workers investigated $H_2^+$ under various conditions, in a variety of studies such as on the exchange energy, on charge exchange, and in calculations of its hyperfine structure. The article by Avanaki, Telnov, and Chu explores $H_2^+$ again, but in a very different context, namely ultrafast pulses of few tens of femtoseconds to use the harmonic generation (HG) process and gain access to the electron dynamics. In fact, near-threshold harmonic spectroscopy can be employed to obtain information on specific structures, in particular bound states and resonances. In this work, the authors study near-threshold harmonics in $H_2^+$, by numerically tuning the electron dynamics via the laser ellipticity, and find that its anomalous ellipticity dependence originates mainly from near-resonant excitation of $\pi_u$ orbitals in $H_2^+$.

**High-resolution spectroscopy of the A$^1\Pi$ (v'=0-10)-X$^1\Sigma^+$(v''=0) bands in $^{13}$C$^{18}$O: Term values, rovibrational oscillator strengths and Hönl-London corrections**
J L Lemaire, M Eidelsberg, A N Heays, L Gavilan, S R Federman, G Stark, J R Lyons, N de Oliveira and D Joyeux

Due to its significant abundance, carbon monoxide (CO) is an important molecule in astronomical environments as varied as interstellar clouds, protoplanetary disks, and planetary and exoplanetary atmospheres to name a few. Dalgarno and co-workers have studied many aspects of CO, such as formation in supernova ejecta and the scattering of CO with other primordial elements, notably H and He. In their article, Lemaire and co-workers have taken advantage of the VUV-FTS end-station of the high-resolution absorption spectroscopy branch of the DESIRS beamline at the SOLEIL Synchrotron to extend their initial work on the A-X bands of $^{12}$C$^{16}$O and $^{13}$C$^{16}$O to the $^{13}$C$^{18}$O isotopologue. The analysis of these spectra allows for precise line assignments, term values, and band-integrated oscillator strengths as well as individual rovibrational oscillator strengths and Hönl-London corrections, and thus provides accurate molecular data to interpret astronomical observations.

**Quadrupole association and dissociation of hydrogen in the early Universe**
Robert C Forrey

The formation and destruction of H$_2$ molecules dictates many of the reactions relevant to the evolution of the chemistry of the early Universe. Dalgarno was a pioneer of these studies, calculating many of the important rate coefficients guiding early Universe chemistry. Forrey explores the formation and dissociation of molecular hydrogen, a homonuclear diatomic molecule without a permanent dipole moment, based on quadrupole transitions. The impact of extremely long-lived resonances is considered for Local Thermodynamic Equilibrium (LTE) and non-LTE conditions prevalent in the early Universe, and it is shown that quadrupole association and dissociation of hydrogen is inefficient, though may have played a role in the early Universe for redshifts $z > 500$.

**H$^-$ photodetachment and radiative attachment for astrophysical applications**
Brendan M McLaughlin, P C Stancil, H R Sadeghpour, and Robert C Forrey

The lightest three-body, two-electron system is H$^-$ and it plays a crucial role in atomic physics. It has been studied extensively both theoretically and experimentally, and is a very important ingredient in astrophysical systems. For example, photodetachment of H$^-$ is a key opacity source in a variety of stellar atmospheres and H$^-$ is essential to understand the recombination era of the early Universe. Here again, the impact of Dalgarno's work has been enormous, as exemplified in the article by McLaughlin and co-workers in which part of the work was initiated by Alex Dalgarno himself. In this article, the authors combine a variety of results to construct the of H$^-$ photodetachment cross section, reliable over a large range of photon energy. They also provide spontaneous and stimulated radiative attachment rate coefficients, and compute photodetachment rates relevant to astronomical conditions.

Some readers may feel some distress about the state of physics, astrophysics, or cosmology due to a perceived increase in speculative and transiently publicized research. A strong antidote to such an ailment is a healthy dose of Dalgarno manuscripts, renowned for their robustness, insight, and conciseness, and their ability to stand the test of time. We hope that the present volume will serve as a testament to Alex Dalgarno's importance and far-reaching impact on modern physics.

**References**

Allison A. C. and Dalgarno A. 1971 *J. Chem. Phys.* **55** 4342-44
Allison D. C. S. and Dalgarno A. 1965 *Proc. Phys. Soc.* **85** 845-849
Amaldi E. and Segre E. 1934 *Nature* **133** 141
Arthurs A. M. and Dalgarno A. 1960 *Proc. R. Soc. London, Ser. A* **256** 540-551
Balakrishnan N., Forrey R. C. and Dalgarno A. 1997a *Chem. Phys. Lett.* **280** 1-4
Balakrishnan N., Kharchenko V., Forrey R. C. and Dalgarno A. 1997b *Chem. Phys. Lett.* **280** 5-9
Chu X. and Dalgarno A. 2004 *J. Chem. Phys.* **121** 4083-88
Côté R. and Dalgarno A. 2000 *Phys. Rev. A* **62** 012709
Dalgarno A. 2000 in *Astrochemistry: From Molecular Clouds to Planetary Systems,* IAU Symposium **197** (Astronomical Society of the Pacific: San Francisco) Y. C. Minh and E. F. van Dishoeck, eds, 1-12
Dalgarno A. 1952 *Proc. Phys. Soc. London, Sec. A* **65** 663-667
Dalgarno A. and McCarroll R. 1956 *Proc. Phys. Soc. London, Sec. A* **237** 383-394
Dalgarno A. and McCarroll R. 1957 *Proc. Phys. Soc. London, Sec. A* **239** 413-419
Dalgarno A. and McCray R. A. 1973 *Astrophys. J.* **181** 95-100
Drake G. W. F. and Dalgarno A. 1969 *Astrophys. J* **157** 459-462
Yan Z.-C., Nörtershaüser W. and Drake G. W. F. 2008 *Phys. Rev. Lett.* **100** 243002
Yan Z.-C., Nörtershaüser W. and Drake G. W. F. 2009 *Phys. Rev. Lett.* **102** 249903 (Erratum)
Field G. B. and Steigman G. 1971 *Astrophys. J.* **166** 59-64
Fermi E. 1934 *Nuovo Cimento* **11**, 157-166
Herbst E. 1981 *Nature* **289** 656-7
Krems R. V. and Dalgarno A. 2004 *J. Chem. Phys.* **120**, 2296-2307
Lepp S. and Dalgarno A. 1988a *Astrophys. J.* **324** 553-556
Lepp S., Dalgarno A., van Dishoeck E.F. and Black J.H. 1988b *Astrophys. J.* **329** 418-424
McCarthy M. C., Gottlieb C. A., Gupta, H. and Thaddeus, P. 2006 *Astrophys. J. Lett.* **652** L141-4.
Rellergert W. G., Sullivan S. T., Kotochigova S., Petrov A., Chen K., Schowalter S. J., and Hudson E. R. 2011 *Phys. Rev. Lett.* **107** 243201
Sun Y. and Dalgarno A., 1994 *Ap. J.* **427** 1053-56
Tajitsu A., Sadakane K., Naito H., Arai A., and Aoki, W. 2015 *Nature* **518**, 381–384
Zygelman B. 1990 *Phys. Rev. Lett.* **64** 256-9

---

[i] For additional autobiographical information about Alexander Dalgarno, the Corresponding Editors recommend *A Serendipitous Journey*, Alexander Dalgarno 2008 *Ann. Rev. Astro. Astrophys.* **46** 1-20. In addition, a video interview conducted by Harry Kreisler of the Institute of International Studies at the Univ. of California Berkeley is available at the links
http://globetrotter.berkeley.edu/people3/Dalgarno/dalgarno-con0.html and
https://www.youtube.com/watch?v=nASubw4qCNM